\documentclass{JAC2003}

\usepackage{graphicx}
\usepackage{booktabs}
\usepackage{multicol}
\usepackage{amsmath}
\usepackage{url}

\begin{document}

\title{Superconducting nano-layer coating without insulator\thanks{The work is supported by JSPS Grant-in-Aid for Young Scientists (B), Number 26800157. }}

\author{
Takayuki Kubo\thanks{kubotaka@post.kek.jp}\\ 
KEK, High Energy Accelerator Research Organization, Tsukuba, Ibaraki 305-0801 Japan
}

\maketitle

\begin{abstract}
The superconducting nano-layer coating without insulator layer is studied. 
The magnetic-field distribution and the forces acting on a vortex are derived. 
Using the derived forces, the vortex-penetration field and the lower critical magnetic field can be discussed. 
The vortex-penetration field is identical with the multilayer coating, 
but the lower critical magnetic field is not. 
Forces acting on a vortex from the boundary of two superconductors play an important role in evaluations of the free energy.  
\end{abstract}

\section{Introduction}

The multilayer coating is an idea for pushing up the field limit of superconducting (SC) accelerating cavity~\cite{gurevich}. 
According to the recent theoretical studies~\cite{kubo, kuboIPAC13, kuboSRF2013, kuboIPAC14}, 
differences among the vortex-penetration fields of semi-infinite SC, SC thin film and multilayer SC are due to those of current densities or slopes of magnetic-field attenuation which are proportional to the force pushing a vortex into SC. 
For the case that a penetration depth of the SC layer is larger than that of the SC substrate, 
the current density or the slope of magnetic-field attenuation in the SC layer is suppressed, 
and the vortex-penetration field of the SC layer is pushed up.

\begin{figure}[*t]
   \begin{center}
   \includegraphics[width=0.8\linewidth]{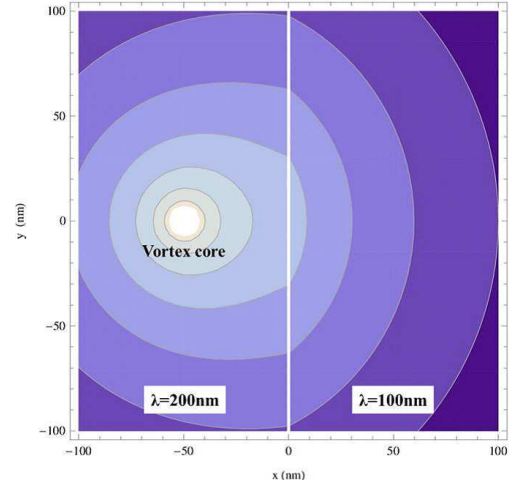}
   \end{center}\vspace{-0.2cm}
   \caption{
A contour plot of the magnetic field around a vortex core near a boundary of two semi-infinite SCs. 
The boundary is indicated by a vertical white-line on $x=0$. 
Penetration depths of left and right SCs are $200\,{\rm nm}$ and $100\,{\rm nm}$, respectively.  
   }\label{fig1}
\end{figure}

A combination of SC materials with different penetration depths is essential, 
but the insulator layer seems to be unnecessary. 
In this paper, the SC nano-layer coating without insulator layer is studied. 
First the magnetic-field distribution is computed. 
Then a set of forces acting on a vortex is derived, 
which is different from that of the multilayer coating model, 
because of the existence of the boundary of two SCs as shown in Fig.~\ref{fig1}. 
By using the forces, the vortex-penetration field and the lower critical magnetic field are discussed.

%

\section{Magnetic-field distribution}

Let us consider a model with an SC layer with a penetration depth $\lambda_1$ and a coherence length $\xi_1$ formed on an SC substrate with a penetration depth $\lambda_2$ and a coherence length $\xi_2$. 
All layers are parallel to the $y$-$z$ plane and then perpendicular to the $x$-axis. 
The magnetic field is applied parallel to the layers. 
We assume the SC layer thickness $d$ is larger than their coherence lengths ($d \gg \xi_1,\, \xi_2$). 
The proximity effect between the two SCs. is neglected.

The variational approach is an easy way to understand how the magnetic-field and the current density  distribute in the system. 
Defining magnetic-fields in the SC layer and SC substrate as $B_1(x)$ and $B_2(x)$, respectively, 
current densities in the SC layer and SC substrate are given by $J_1(x)=-\mu_0^{-1}dB_1/dx$ and $J_2(x)=-\mu_0^{-1}dB_2/dx$, respectively. 
Then the total energy of the system as a functional of $B_k$ ($k=1,2$) is given by 
\begin{eqnarray}
E[B_k] \!=\!\! \int \!\! dydz \sum_{k=1}^{2} \int_{L_{k-1}}^{L_{k}}\!\!\!\!\!\! dx \Bigl( \frac{B_k^2}{2\mu_0} + \frac{\mu_0}{2}\lambda_k^2 J_k^2 \Bigr) \,, 
\end{eqnarray}
where $L_0 \equiv 0$, $L_1 \equiv d$, $L_2 \equiv \infty$. 
Variation of the above functional vanishes when conditions 
\begin{eqnarray}
\frac{d^2 B_k}{dx^2} &=& \frac{B_k}{\lambda_k^2}  \,, \label{eq:london} \\
B_1(d) &=& B_{2}(d)  \,, \label{eq:boundaryB} \\
\lambda_1^2 \frac{dB_1}{dx}\biggr|_{d} &=& \lambda_{2}^2 \frac{dB_{2}}{dx}\biggr|_{d} \,, \label{eq:boundaryA}
\end{eqnarray}
are satisfied. 
Eq.~(\ref{eq:london}) is the London equation in each SC layer. 
Eq.~(\ref{eq:boundaryB}) and (\ref{eq:boundaryA}) are the continuity condition of the magnetic field and the vector potential, respectively. 
The above formalism can be easily generalized to an $n$ SC-layer system ($n\ge 3$).

\begin{figure}[*t]
   \begin{center}
   \includegraphics[width=0.9\linewidth]{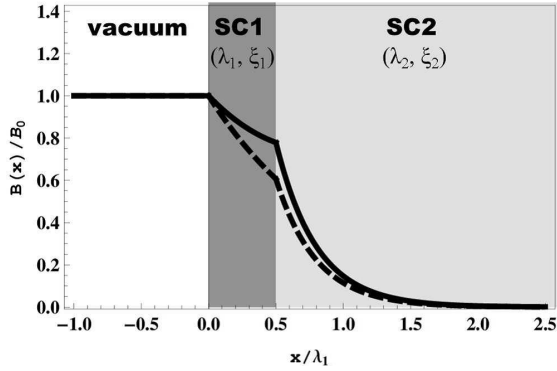}
   \end{center}\vspace{-0.2cm}
   \caption{
The magnetic-field distribution in the SC nano-layer coating without insulator layer. 
A solid curve represent the magnetic-field distribution given by Eq.~(\ref{B1}) and (\ref{B2}). 
A dotted curve represents a naive exponential decay for comparison with the correct curve. 
   }\label{fig2}
\end{figure}

Solving the above equations, we obtain 
\begin{eqnarray}
B_1(x) &=& 
B_0
\frac{ \cosh \frac{d-x}{\lambda_1} + \frac{\lambda_2}{\lambda_1} \sinh \frac{d-x}{\lambda_1}}
     { \cosh \frac{d}{\lambda_1} + \frac{\lambda_2}{\lambda_1} \sinh \frac{d}{\lambda_1}} \,  ,  
\label{B1} \\
B_2(x) &=& B_1(d) e^{-\frac{x-d}{\lambda_2}} \, , 
\label{B2}
\end{eqnarray}
where $B_0 \equiv B_1(0)$. 
The result is identical with the magnetic-field distribution of the multilayer SC with $d_{\mathcal I}\to 0$. 
Fig.~\ref{fig2} shows how a magnetic field attenuates in the system. 
The slope of  attenuation is suppressed in common with the multilayer SC with insulator layer. 

%

\section{Forces acting on a vortex} 

Suppose there exist a vortex with the flux quantum $\phi_0=2.07\times10^{-15}{\rm Wb}$ parallel to ${\bf \hat{z}}$ at $x=x_0$, where materials of the SC layer and the SC substrate are assumed to be extreme type II SCs, namely, $\lambda_k\gg \xi_k$ ($k=1,2$). 
This vortex feels three distinct forces ${\bf f}_{\rm S}(x_0)$, ${\bf f}_{\rm B}(x_0)$ and ${\bf f}_{\rm M}(x_0)$, 
where ${\bf f}_{\rm S}$ is a force from the surface, 
${\bf f}_{\rm B}$ is that from the boundary of two SCs, and
${\bf f}_{\rm M}$ is that from a Meissner current due to an applied magnetic-field. 

\subsection{Vortex in the SC Layer ($\xi_1 \le x_0 \le d-\xi_1$)} 

Let us consider the case that a vortex is located at the region $\xi_1 \le x_0 \le d-\xi_1$. 
The forces, ${\bf f}_{\rm S}$ and ${\bf f}_{\rm B}$, can be evaluated by introducing image vortices to satisfy the boundary conditions (BCs): the zero current normal to the surface and the continuity condition of the vector potential at the boundary of two SCs. 
For simplicity we consider only two dominant images, 
an image antivortex with flux $\phi_0$ at $x=-x_0$ and an image vortex with flux 
\begin{eqnarray}
\phi_1 \equiv \frac{\lambda_1^2-\lambda_2^2}{\lambda_1^2 + \lambda_2^2}\phi_0
\end{eqnarray}
at $x=2d-x_0$. 
The former and the latter correspond to BCs at the surface and the boundary, respectively. 
Then we find 
\begin{eqnarray}
{\bf f}_{\rm S}(x_0) 
&=& - \frac{\phi_0^2}{4\pi\mu_0 \lambda_1^2  x_0} {\bf \hat{x}} \, , \label{fS1} \\
{\bf f}_{\rm B}(x_0) 
&=& -\frac{\phi_0 \phi_1}{4\pi\mu_0 \lambda_1^2 (d-x_0)} {\bf \hat{x}} \, , \label{fB1}
\end{eqnarray}
where ${\bf \hat{x}}=(1,0,0)$. 
The surface attracts the vortex and prevent the vortex penetration, and 
the boundary pushes the vortex to the direction of the SC material with a larger penetration depth, 
and thus prevent the vortex penetration if $\lambda_1 > \lambda_2$. 
The force, ${\bf f}_{\bf M}$, is obtained by evaluating the Meissner-current density at the vortex position $x= x_0$(see Ref.~\cite{kuboSRF2013, kuboIPAC14}) and is given by
\begin{eqnarray}
{\bf f}_{\rm M}(x_0)
=\frac{B_0\phi_0}{\mu_0 \lambda_1}
\frac{ \sinh \frac{d-x_0}{\lambda_1} + \frac{\lambda_2}{\lambda_1} \cosh \frac{d-x_0}{\lambda_1}}
{\cosh \frac{d}{\lambda_1} + \frac{\lambda_2}{\lambda_1}\sinh \frac{d}{\lambda_1}} {\bf \hat{x}} \,, \label{fM1}
\end{eqnarray}
which always pushes the vortex to the inside. 

\subsection{Vortex in the SC Substrate ($d + \xi_2 < x_0 < \infty$)} 

For the case that a vortex is at $d + \xi_2 < x_0 < \infty$, in order to satisfy BCs at the surface and the boundary, we introduce following two images: 
an image antivortex with flux 
\begin{eqnarray}
\phi_2 \equiv \frac{2\lambda_1^2}{\lambda_1^2 + \lambda_2^2} \frac{2\lambda_2^2}{\lambda_1^2 + \lambda_2^2}\phi_0 
\end{eqnarray}
at $x=-x_0$ and an image antivortex with flux $\phi_1$ at $x=2d-x_0$. 
Then we find
\begin{eqnarray}
{\bf f}_{\rm S}(x_0) 
&=& - \frac{\phi_0 \phi_2}{4\pi\mu_0 \lambda_2^2 x_0} {\bf \hat{x}} \, , \label{fS2} \\
{\bf f}_{\rm B}(x_0) 
&=& -\frac{\phi_0 \phi_1}{4\pi\mu_0 \lambda_2^2 (x_0-d)} {\bf \hat{x}} \, . \label{fB2}
\end{eqnarray}
The surface and the boundary, if $\lambda_1 > \lambda_2$, prevent the vortex penetration as the above. 
${\bf f}_{\bf M}$ can be obtained in much the same way as before: 
\begin{eqnarray}
{\bf f}_{\rm M}(x_0)
=\frac{B_1(d)\phi_0}{\mu_0 \lambda_2} e^{-\frac{x_0-d}{\lambda_2}}
{\bf \hat{x}} \,, \label{fM2}
\end{eqnarray}
which pushes the vortex to the inside.

\section{Vortex-penetration field} 

\begin{figure}[*t]
   \begin{center}
   \includegraphics[width=0.9\linewidth]{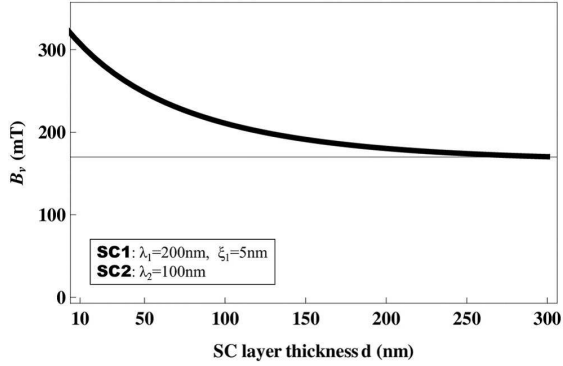}
   \end{center}\vspace{-0.2cm}
   \caption{
The vortex-penetration field of the SC nano-layer coating without insulator layer as a function of SC layer thickness.   
   }\label{fig3}
\end{figure}

The vortex-penetration field, at which the Bean-Livingston barrier disappears, 
can be obtained by balancing the forces acting on a vortex at the surface ($x_0=\xi_1$). 
Substituting $x_0=\xi_1$ into Eq.~(\ref{fS1}), (\ref{fB1}) and (\ref{fM1}), and 
imposing the condition ${\bf f}_{\rm S}(\xi_1)+{\bf f}_{\rm B}(\xi_1) + {\bf f}_{\rm M}(\xi_1)= {\bf 0}$, 
we obtain
\begin{eqnarray}
B_{v}
= 
\frac{\phi_0}{4\pi \lambda_1 \xi_1} 
\frac{\cosh\frac{d}{\lambda_1} \!+\! \frac{\lambda_2}{\lambda_1} \sinh\frac{d}{\lambda_1}}
       {\sinh\frac{d}{\lambda_1} \!+\! \frac{\lambda_2}{\lambda_1} \cosh\frac{d}{\lambda_1} }\,,
\label{eq:BvML}
\end{eqnarray}
where the contribution from $|{\bf f}_{\rm B}(\xi_1)| \,(\ll |{\bf f}_{\rm S}(\xi_1)|)$ is neglected. 
For a thin SC layer, the vortex-penetration field is enhanced up to 
$B_{v}|_{d\ll\lambda_1}\simeq (\phi_0/4\pi \lambda_1 \xi_1) (\lambda_1/\lambda_2)$. 
As expected, the SC nano-layer coating model without insulator layer also enhance the vortex-penetration field as well as the top SC layer of the multilayer SC (see Fig.~\ref{fig3}). 

%

\section{lower critical magnetic field} 

The lower critical magnetic field $B_{c1}$ is the applied magnetic-field at which a stable position of the vortex changes from the surface to the inside of the SC. 
In other words, the applied magnetic-field at which the free-energy difference between the system with a vortex at the surface and that with a vortex at the stable position inside the SC vanishes.
The free-energy difference between different vortex positions can be evaluated by integrating forces acting on the vortex.\footnote{For example, let us derive $B_{c1}$ of the SC thin film with a penetration depth $\lambda$, a coherence length $\xi$ and a thickness $d\,(\xi \ll d \ll \lambda)$. 
The forces acting on a vortex at $x_0$ from both the surfaces is approximately given by contributions from two images, $f_S \simeq -(\phi_0^2/4\pi \mu_0 \lambda^2) [1/x_0 - 1/(d-x_0)]$. 
Note that the second term, which can be neglected for a calculation of the vortex-penetration field~\cite{kuboSRF2013, kuboIPAC14}, is necessary to evaluate $B_{c1}$.  
The force due to the Meissner current~\cite{kuboSRF2013, kuboIPAC14} is given by $f_M = (B_0 \phi_0/\mu_0 \lambda) [\sinh (d/2\lambda - x_0/\lambda)/\cosh(d/2\lambda) ]$. 
The stable position of the vortex is  $x_0=d/2$. 
Then the energy difference between the system with a vortex at the surface $x_0=\xi$ and that with a vortex at $x_0 =d/2$ is given by
\begin{eqnarray}
\Delta G 
= -\int_{\xi}^{\frac{d}{2}}dx_0 (f_S + f_M)
\simeq \frac{\phi_0^2}{4\pi \mu_0 \lambda^2}\ln \frac{d}{\xi} - \frac{B_0 \phi_0}{\mu_0} \frac{d^2}{8\lambda^2}\,.
\nonumber
\end{eqnarray}
The condition $\Delta G = 0$ yields 
\begin{eqnarray}
B_{c1} \simeq \frac{2\phi_0}{\pi d^2} \ln \frac{d}{\xi} \,, \nonumber
\end{eqnarray}
which corresponds to the result given in Ref.~\cite{stejic}.}

\begin{figure}[*t]
   \begin{center}
   \includegraphics[width=0.9\linewidth]{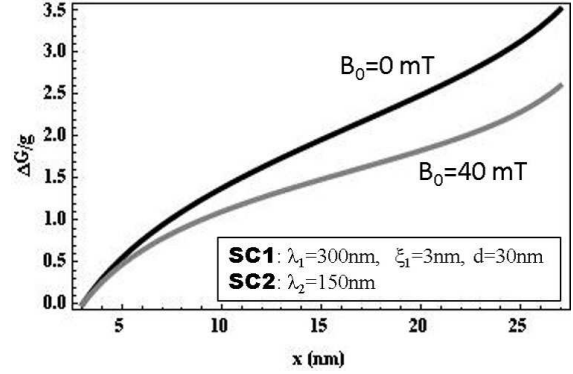}
   \end{center}\vspace{-0.2cm}
   \caption{
The free energy difference (in an unit of $g\equiv \phi_0^2/4\pi\mu_0\lambda_1^2$) between a system with a vortex at the surface and that with a vortex inside the SC layer with thickness $30\,{\rm nm}$. 
   }\label{fig4}
\end{figure}

We show only a part of the results. 
Fig.~\ref{fig4} shows the free energy difference inside the SC layer for the case that zero or small magnetic field is applied. 
Since both the surface and the boundary pushes a vortex to the surface, 
the free energy increases as a vortex approaches the boundary. 
Thus the boundary is not necessarily a stable position of the vortex\footnote{It should be noted that the boundary can be a metastable position due to the energy gap at the boundary. 
For simplicity, let us consider the case that the SC layer is replaced with a semi-infinite SC, 
namely, a system that consists of two semi-infinite SCs~\cite{gap} with a boundary at $x=d$. 
In this case, the free energy at $x=d-\xi_1$ and  $x=d+\xi_2$ are given by $G_1=\epsilon_1 + (\phi_0\phi_1/4\pi\mu_0\lambda_1^2)\ln\kappa_1$ and $G_2=\epsilon_2 - (\phi_0\phi_1/4\pi\mu_0\lambda_2^2)\ln\kappa_2$, respectively, 
where $\epsilon_k=(\phi_0^2/4\pi\mu_0\lambda_k^2)\ln\kappa_k$ ($k=1,2$) is the free energy of an isolated vortex in each SC. 
Thus the energy gap at the boundary is given by 
\begin{eqnarray}
\Delta G_{\rm gap} = G_1-G_2 = \frac{\phi_0^2}{2\pi\mu_0(\lambda_1^2+\lambda_2^2)}\ln\frac{\kappa_1}{\kappa_2} \,,  \nonumber
\end{eqnarray}
which can make a valley of the free energy. 
} 
in contrast to the insulator-side surface of the multilayer SC.\footnote{The insulator-side surface of the multilayer SC attracts a vortex, and the free energy decreases as a vortex approaches the insulator-side surface. 
As a result the insulator-side surface becomes a stable position of the vortex even if the applied magnetic-field is very small~\cite{posen}.}
To know the stable position and evaluate the lower critical magnetic field, 
the free energy calculations of entire region including an energy gap at the boundary are necessary~\cite{kuboElsewhere}.

\section{Summary}

We studied the SC nano-layer coating without insulator. 
\begin{Itemize}
\item{
The magnetic-field distribution is identical with that of the multilayer SC with an insulator layer thickness $d_{\mathcal I}\to 0$. 
}
\item{
The boundary of two SCs introduces a force that pushes a vortex to the direction of material with larger penetration depth. 
}
\item{
The vortex-penetration field is identical with that of the multilayer SC with $d_{\mathcal I}\to 0$,  
because the difference between the systems with and without insulator is negligible at the surface.
}
\item{
The boundary is not necessarily a stable position of the vortex, in contrast to the insulator-side surface of the multilayer SC. 
To know the stable position and evaluate $B_{c1}$, the free energy calculations of entire region including an energy gap at the boundary are necessary. 
}
\end{Itemize}
The SC nano-layer coating without insulator might be a possible idea to enhance the vortex-penetration field without sacrificing $B_{c1}$. 
Results and detailed discussions on $B_{c1}$ will be presented~\cite{kuboElsewhere}.

\end{document}